\documentclass[journal]{IEEEtran}
\usepackage{graphicx}
\usepackage{dcolumn}
\usepackage{bm}
\usepackage{xspace}
\usepackage{amsfonts}
\usepackage{amsmath}
\usepackage{amssymb}
\usepackage{epsfig}
\usepackage{pslatex}
\usepackage{dblfloatfix}
\usepackage{cite}
\usepackage{textcomp}

\newcommand{\mum}{\text{\textmu m}}

\newcommand{\SiN}{\text{Si}_\text{3}\text{N}_\text{4}}

\begin{document}
%
\title{Si$_3$N$_4$ nanobeam optomechanical crystals}
%
%
%

\author{Karen E. Grutter,
	Marcelo~Davan\c co,
	and Kartik~Srinivasan%
\thanks{K.E. Grutter, M. Davan\c co, and K.Srinivasan are with the Center for Nanoscale Science and Technology, National Institute of Standards and Technology, Gaithersburg, MD 20899-6203, USA (e-mail: karen.grutter@nist.gov; marcelo.davanco@nist.gov, kartik.srinivasan@nist.gov)}}
\IEEEspecialpapernotice{(Invited Paper)}

%



\maketitle

\begin{abstract}
The development of $\SiN$ nanobeam optomechanical crystals is reviewed.  These structures consist of a 350~nm thick, 700~nm wide doubly-clamped $\SiN$ nanobeam that is periodically patterned with an array of air holes to which a defect region is introduced.  The periodic patterning simultaneously creates a photonic bandgap for 980~nm band photons and a phononic bandgap for 4~GHz phonons, with the defect region serving to co-localize optical and mechanical modes within their respective bandgaps.  These optical and mechanical modes interact dispersively with a coupling rate $g_{0}/2\pi\approx$100~kHz, which describes the shift in cavity mode optical frequency due to the zero-point motion of the mechanical mode. Optical sidebands generated by interaction with the mechanical mode lie outside of the optical cavity linewidth, enabling possible use of this system in applications requiring sideband-resolved operation. Along with a review of the basic device design, fabrication, and measurement procedures, we present new results on improved optical quality factors (up to $4\times10^5$) through optimized lithography, measurements of devices after HF acid surface treatment, and temperature dependent measurements of mechanical damping between 6~K and 300~K. A frequency-mechanical quality factor product $\left(f{\times}Q_m\right)$ as high as $\approx2.6\times10^{13}$~Hz is measured.
\end{abstract}


%
\IEEEpeerreviewmaketitle

\section{Introduction}
\label{sec:Intro}
\IEEEPARstart{C}{avity} optomechanical systems are being developed for a wide variety of purposes, including applications in sensing and metrology~\cite{ref:Aspelmeyer_Kippenberg_Marquardt_Review,ref:Schliesser_resolved_sideband2,ref:Teufel_nanomechanical_motion_Heisenberg,ref:Srinivasan18,ref:Krause_Painter_accelerometer,ref:Forstner_Bowen_magnetometer}, signal transduction and wavelength conversion using the radiation pressure coupling between optics and mechanics~\cite{ref:Hill_Painter_WLC_Nat_Comm,ref:Dong_Wang_Science_dark_mode,ref:Liu_yuxiang_wlc,ref:Bochmann_Cleland_AlN_uwave_optomechanics,ref:Bagci_Polzik_radio_wave_nanomechanics,ref:Andrews_Lehnert_uwave_to_optical}, and the generation of non-classical states of light~\cite{ref:Brooks_Stamper_Kurn_squeezing,ref:safavi-naeini_squeezing,ref:Purdy_Regal_squeezing}.  Milestone experimental demonstrations of ground state cooling~\cite{ref:Teufel_ground_state,ref:Chan_Painter_ground_state}, parametrically-driven normal mode splitting~\cite{ref:Groblacher_Aspelmeyer_strong_coupling,ref:Verhagen_Kippenberg_Nature}, and coherent energy transfer between the optical and mechanical domains~\cite{ref:Fiore_Wang_PRL,ref:Palomaki_Lehnert_state_transfer} have inspired numerous theoretical proposals that use coupled photons and phonons for applications in areas such as quantum information science~\cite{ref:Stannigel_Lukin_prl,ref:Stannigel_Rabl_prl,ref:Ludwig_Marquardt,ref:Rips_Hartmann_nanomechanical_qubits,ref:Borkje_nonlinear_cavity_OM}.

The optomechanical crystal platform developed by Painter and colleagues~\cite{ref:eichenfield2,ref:Safavi-Naeini1} seeks to combine localized and interacting photons and phonons in a cavity optomechanical system with propagating photons and phonons in bus waveguides that route signals on the chip~\cite{ref:safavi-naeini3}.  Both one-dimensional~\cite{ref:eichenfield2,ref:safavi-naeini4,ref:Chan_Painter_ground_state} and two-dimensional~\cite{ref:Alegre,ref:safavi_naeini_snowflake_prl} systems have been developed in silicon-on-insulator, with demonstrations of a large optomechanical coupling strength $g_{0}/2\pi\approx1$~MHz (which corresponds to the frequency shift of the optical mode due to the zero point motion of the mechanical mode)~\cite{ref:chan_optimized_OMC}, high optical quality factor $Q_{o}\gtrsim10^6$, and high mechanical quality factor $Q_{m}\approx10^6$ at 10~K~\cite{ref:chan_optimized_OMC} and $Q_{m}\approx10^7$ at 300~mK~\cite{ref:Meenehan_Painter_pra}.  The availability of mature planar silicon fabrication technology also enables, among other things, potential electrostatic integration for biasing, readout, and control of the devices.

Despite the many advantages of working in silicon-on-insulator, there are good reasons to investigate optomechanical crystals in other material systems. For example, two-photon absorption in silicon at 1550~nm~\cite{ref:Lin_Painter_Agrawal} and the accompanying generation of free carriers ultimately limits the number of photons that can be injected into a silicon optomechanical resonator without degrading its optical quality factor, mechanical quality factor, or both.  For example, in Ref.~\cite{ref:Chan_Painter_ground_state}, two-photon absorption effects are observed at an intracavity photon population of a few hundred photons, which limits how low a phonon occupancy can be achieved in laser cooling experiments. Incorporation of a p-n junction to sweep out free carriers~\cite{ref:Rong_Si_Raman_laser} and operation at wavelengths above 2000~nm~\cite{ref:Liu_Green,ref:Zlatanovic} have been used to mitigate two-photon absorption and free-carrier generation effects in lasing and nonlinear optics applications, but these non-trivial steps have not yet been demonstrated in the context of cavity optomechanics.  Operation in silicon also restricts the range of available wavelengths over which the optomechanical system may operate to the near-infrared region $>1000$~nm.  Access to shorter wavelengths may be important in a number of scenarios, including stabilization of cavity optomechanical systems to the transitions of alkali atoms and interfacing cavity optomechanical devices with quantum optical systems based on semiconductor quantum dots, defect color centers in crystals, and trapped ions and atoms.

\begin{figure*}[!t]
\centerline{\includegraphics[width=0.9\linewidth]{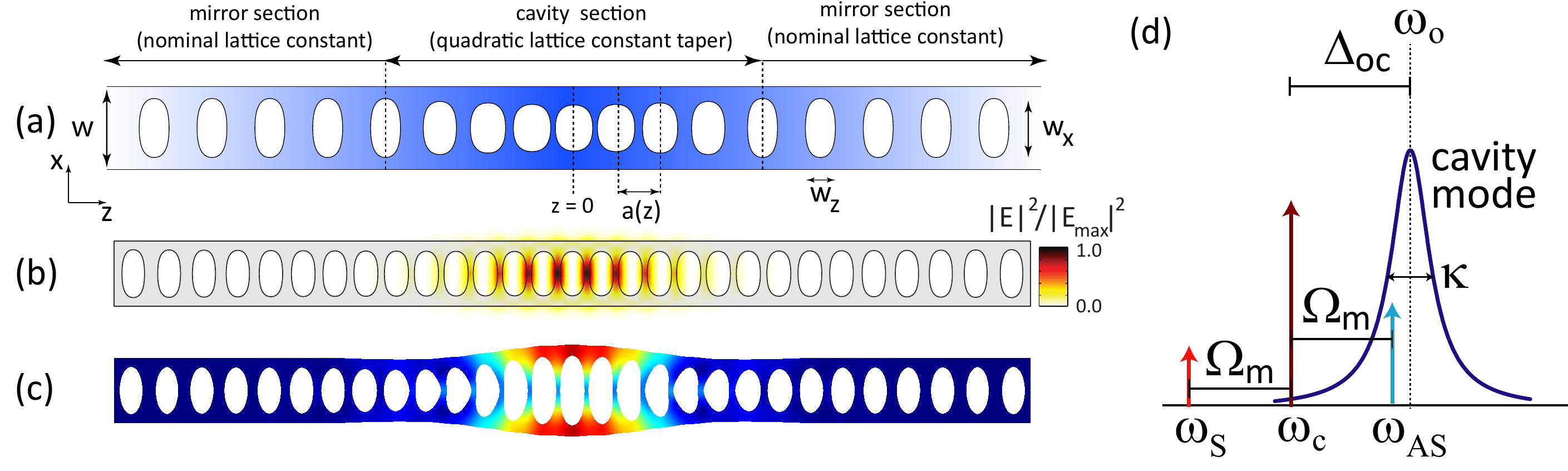}}\caption{(a) Top view schematic of optomechanical crystal geometry. (b) First-order optical resonance at $\lambda \approx980$~nm. (c) First-order breathing mechanical mode at $f_m \approx$3.5~GHz. (d) Frequency domain schematic of a sideband-resolved cavity optomechanical system.  The mechanical resonator generates Stokes ($\omega_{\text{S}}$) and anti-Stokes ($\omega_{\text{AS}}$) sidebands around an input optical beam ($\omega_{c}$).  When the optical cavity mode linewidth $\kappa$ is sufficiently narrow with respect to the mechanical resonator frequency $\Omega_{m}$, it can preferentially select and enhance the generation of one of the two sidebands.} \label{Fig_1}
\end{figure*}

With that motivation, we have recently investigated one-dimensional optomechanical crystals in stoichiometric silicon nitride ($\SiN$)~\cite{ref:Davanco_nanobeam_OMC} (Fig.~\ref{Fig_1}), demonstrating sideband-resolved devices in which the mechanical mode frequency $\Omega_{m}$ exceeds the cavity mode optical linewidth $\kappa$, and observing phenomena that take advantage of the ability of the optical cavity to preferentially select one optical sideband scattered by the mechanical resonator, such as optomechanically-mediated electromagnetically induced transparency~\cite{ref:Weis_Kippenberg,ref:safavi-naeini4}.  In the first part of this paper, we review these recent results and provide additional experimental details not present in our earlier work. We then describe new device fabrication and measurements aimed at further exploring the potential of these devices. In particular, we describe improvements to the fabrication process (in particular, electron beam lithography) that have increased the optical quality factor of the devices by a factor of $\approx$~4 (up to $4{\times}10^5$).  We also discuss the use of weak HF acid etching, meant to remove a thin surface layer of material that may be damaged from the plasma dry etching process used to pattern the nanobeams, as a means to improve device performance.  Finally, we present temperature-dependent measurements of mechanical damping of the $\approx3.8$~GHz breathing mode between 6~K and 300~K.

\section{Overview}
\label{sec:overview}

\subsection{Relationship to Other $\SiN$ Cavity Optomechanical Systems}
Silicon nitride has been used by many researchers to demonstrate high quality factor mechanical resonators ($Q_{m}\gtrsim10^6$) in membrane~\cite{ref:Zwickl_Harris} and doubly-clamped beam geometries~\cite{ref:Verbridge}, with frequencies ranging from hundreds of kHz to tens of MHz. Typically researchers use stoichiometric $\SiN$ grown by low-pressure chemical-vapor deposition, though low-stress, non-stoichiometric SiN has also been used.  Si$_3$N$_4$ resonators have been used in many cavity optomechanics experiments; for example, the $\SiN$ membranes have been used as a dispersive element within a Fabry-Perot cavity~\cite{ref:Thompson_Harris,ref:Purdy_Regal_squeezing,ref:Kemiktarak_Lawall}.  Within integrated on-chip geometries, researchers have developed $\SiN$ cavity optomechanical systems using nanobeams evanescently coupled to whispering gallery mode cavities~\cite{ref:Anetsberger_near_field} and acting as parts of on-chip interferometers~\cite{ref:Fong_Tang}.  $\SiN$ has also been used in double microdisk geometries, where the optical modes supported by vertically-coupled $\SiN$ microdisks are coupled to the out-of-plane flapping motion of the two disks~\cite{ref:Wiederhecker_Lipson}. Similarly, laterally-coupled photonic crystal nanobeam cavities in $\SiN$ have been demonstrated, where the optical mode supported by these beams is coupled to the antisymmetric in-plane mechanical motion of the beams~\cite{ref:eichenfield1,ref:Krause_Painter_accelerometer}.  Optomechanical oscillators based on $\SiN$ whispering-gallery-mode resonators and operating at $\approx~$50~MHz mechanical resonant frequencies have also recently been demonstrated~\cite{ref:Beyazoglu_OMO,tallur2011monolithic}.

When an optical field is directed onto a mechanical resonator, its motion creates high- and low-frequency sidebands around the optical field frequency due to the Doppler effect (this is essentially the same process as Raman scattering).  In a cavity optomechanical system, the optical cavity has an associated density of electromagnetic states, and depending on the detuning of the probe laser, the optical cavity can preferentially enhance and select the creation of one sideband with respect to the other.  The degree of asymmetry is determined by the width of the optical cavity mode ($\kappa$) compared to the frequency of the mechanical resonator ($\Omega_{m}$), as the generated sidebands are separated from the input optical beam by $\Omega_{m}$ (Fig.~\ref{Fig_1}(d)).  Systems for which $\Omega_{m}>\kappa$ are said to be sideband-resolved~\cite{ref:Kippenerbg_Vahala_Science}, an important criterion for applications such as laser cooling, where the level of sideband resolution sets an ultimate limit on the achievable phonon occupancy of the mechanical resonator~\cite{ref:Wilson-Rae_Kippenberg_OM_cooling,ref:Marquardt_Girvin_OM_cooling}.  The chip-based $\SiN$ systems discussed above are not in the sideband-resolved regime, limiting their potential in such applications.

The $\SiN$ nanobeam optomechanical crystals first presented in Ref.~\cite{ref:Davanco_nanobeam_OMC} and developed further here are distinguished by operating in the sideband-resolved regime (Fig.~\ref{Fig_1}).  This is primarily accomplished through the use of highly localized mechanical modes whose frequencies are in the GHz range (2-3 orders of magnitude larger than the mechanical frequencies of the devices referenced above), exceeding optical cavity loss rates that are usually in the hundreds of MHz range.  We note that even in state-of-the-art $\SiN$ nanophotonic devices, optical quality factors are typically in the $10^6$ range ($\kappa/2\pi\gtrsim~$100~MHz)~\cite{ref:Barclay8,ref:Gondarenko,ref:Hosseni_Adibi}, though higher $Q_{o}$ values of $2{\times}10^7$~\cite{ref:Li_Adibi_high_Q_SiN} and $8{\times}10^7$~\cite{ref:Spencer_Bowers_high_Q} have recently been achieved.  These recent demonstrations have come at the expense of optical field confinement; in the former, 240~$\mum$ radius microrings with a 400~nm $\SiN$ layer were studied, while in the latter, a thin $\SiN$ layer of 40~nm is used to create a delocalized optical mode that primarily sits in the surrounding SiO$_2$, restricting ring radii to several mm. On the other hand, small diameter (10~$\mum$) $\SiN$ microdisks have recently been demonstrated and used in cavity optomechanical wavelength conversion experiments~\cite{ref:Liu_yuxiang_wlc}, where sideband-resolved operation was achieved with a 625~MHz mechanical mode, exceeding the cavity mode optical linewidth $\kappa/2\pi\approx$~150~MHz.  However, the optomechanical coupling rate in that system is more than one order of magnitude smaller than the rate in the $\SiN$ nanobeam optomechanical crystals described below.

\subsection{Optomechanical Design}

This section contains a brief outline of the design process for optomechanical crystal cavities. Further details are given elsewhere~\cite{ref:Davanco_OMC,ref:Davanco_nanobeam_OMC}.
The optomechanical crystal (OMC) consists of a suspended $\SiN$ nanobeam of thickness $t$ and width $w$,
with an etched array of elliptical holes as illustrated in Fig.~\ref{Fig_1}(a).
In the outer mirror sections, the spacing $a(z)$ between the holes is constant, while within
the cavity section, it varies quadratically from the center
outwards. The cavity section forms a defect in an otherwise perfect 1D photonic bandgap structure, where the optical field can be longitudinally confined (along the $z$ axis). Longitudinal optical confinement is achieved when frequencies allowed to propagate along $z$ in the cavity section fall within the photonic bandgap of the mirror sections. Simultaneously with a photonic bandgap, the 1D photonic crystal structure forming the mirrors also provides a bandgap for mechanical waves---a phononic bandgap---such that the defect in the cavity region can also support localized mechanical resonances~\cite{ref:eichenfield2}. The optomechanical crystal design process consists of creating an appropriate modulated array of holes that maximizes the radiation-pressure interaction between the localized optical and mechanical resonances, by maximizing the spatial overlap between the two; further requirements are that the optical and mechanical quality factors be maximized.

The design procedure starts with a consideration of the optical resonances. Photonic bandgaps can be calculated with a variety of methods; in our case, the plane-wave expansion method was used. The lattice spacing $a$ at the cavity center is initially chosen such that a specific target wavelength is allowed to propagate, at the very edge of the photonic bandgap. The lattice spacing is then allowed to vary from the value at the center, so that the target frequency falls within the bandgap of the crystal in the mirror sections. The lattice spacing variation must be carefully chosen in order to minimize coupling to free-space radiation, so that high optical quality factors ($Q_o>10^6$) can be achieved~\cite{ref:Quan1,ref:Davanco_nanobeam_OMC}. The modulated hole array can be regarded as a 1D distributed feedback mirror with a variable mirror strength. Linear mirror strength profiles have been shown to produce optical modes with
reduced spatial harmonics above the light line, which leads to reduced power leakage into the air---and thus higher quality
factors~\cite{ref:Quan1,ref:Srinivasan1}. This procedure has been used to produce theoretical $Q_o>10^6$ at various wavelengths (in the present work, near 980~nm).

As discussed above, the photonic bandgap structure also acts as a phononic bandgap structure. In particular, photonic crystal cavities optimized with the procedure above already display confined mechanical resonances that are co-located with the optical mode, but with sub-optimal spatial overlap, and thus lower optomechanical interaction strength. The optomechanical interaction strength is quantified by the optomechanical coupling rate $g_0$, which is the coupling rate between single photons and phonons in the cavity. This parameter can be calculated with a perturbative expression involving an overlap integral between the optical and mechanical displacement fields~\cite{ref:eichenfield2,ref:Davanco_OMC}. Optical and mechanical cavity modes can be calculated using a variety of computational methods; in our case, the finite element method was used.

In order to optimize the localization of the mechanical resonance, the aspect ratio $w_x/w_z$ of the lattice holes is allowed to vary along the nanobeam, with constant $w_x\cdot w_z$. For sufficiently small aspect ratio excursions, the optical modes are only perturbatively affected, and optical $Q$s are not significantly degraded. At this point, a nonlinear search routine can be applied to maximize the optomechanical coupling rate $g_0$ and optical quality factor through small adjustments to both the lattice constant and hole aspect ratio profiles along the nanobeam. Figs.~\ref{Fig_1}(b) and (c) show the co-localized, confined optical and breathing mechanical modes resulting from one such optimization process, for a $\SiN$ nanobeam OMC. The optical mode is at a $971$~nm wavelength and has $Q_o>10^6$. The breathing mechanical mode is such that the nanobeam sidewalls expand and contract laterally (in the $x$ axis) in a breathing fashion, with a frequency of $\approx3.5$~GHz.

We point out that in this work, contributions from stress-induced changes of the refractive index (photoelastic effect) were not considered in the optimization process. This contribution has been shown to be significant in crystalline silicon ~\cite{ref:chan_optimized_OMC,ref:rakich_PRX} and GaAs~\cite{ref:KCB_gaas_OM_optica} nanostructures, for which the photoelastic tensor has been measured \cite{biegelsen1974photoelastic,dixon1967photoelastic}. However, we have not found analogous measurements for the photoelastic tensor in $\SiN$.

\subsection{Optomechanical Spectroscopy}
\label{sec:OM_spectroscopy}

\begin{figure*}[!t]
\centering
\includegraphics[width=0.8\linewidth]{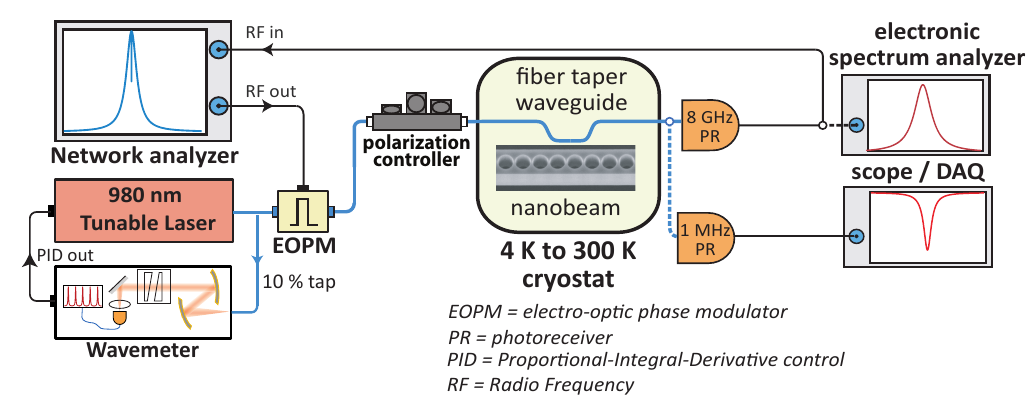}
 \caption{Experimental setup for optical and mechanical mode characterization. Devices are tested in either an ambient, plexiglass-enclosed setup or a 4~K to 300~K cryogen-free cryostat, with light coupled into and out of the nanobeam optomechanical crystals using a fiber taper waveguide.  A 980~nm band external cavity tunable diode laser is used for swept wavelength spectroscopy of optical cavity modes, with the transmission signal detected on a 1~MHz photoreceiver.  Mechanical mode spectroscopy is performed by fixing the laser on the shoulder of the optical cavity mode and detecting the transmission signal on an 8~GHz photoreceiver, the output of which is sent to a real-time electronic spectrum analyzer.  For sideband spectroscopy/electromagnetically induced transparency and absorption measurements, the laser wavelength is fixed at a frequency $\omega_{c}$ that is detuned from the optical cavity mode by $\Delta_{oc}=\omega_{o}-\omega_{c}$.  Modulation sidebands are created with an electro-optic phase modulator (EOPM) driven by a vector network analyzer. The vector network analyzer also demodulates the electrical signal from the 8~GHz photoreceiver, so that as the modulation frequency is swept, spectra of the sideband probe transmission under application of a fixed frequency control field are generated. A wavemeter is used to monitor the 980~nm laser wavelength and if necessary, provide feedback to the laser to stabilize its wavelength.} \label{fig:Figure1}
\end{figure*}

The experimental setup used to measure the $\SiN$ nanobeam optomechanical crystals is shown in Fig.~\ref{fig:Figure1}, and was previously described in \cite{ref:Davanco_nanobeam_OMC,ref:Liu_yuxiang_wlc}.  Light from a 980~nm external cavity tunable diode laser is coupled in and out of the devices using an optical fiber taper waveguide (FTW) with a diameter of $\approx$~1~$\mum$. The fiber taper is typically positioned several hundred nanometers to the side of the nanobeam, to limit parasitic loading of the optical cavity. Optical cavity mode spectroscopy is performed by sweeping the laser wavelength and detecting the transmitted signal through the fiber taper on a 1~MHz photoreceiver, with the laser sweep span calibrated by a wavemeter.  Mechanical mode spectroscopy is performed by fixing the laser wavelength on the side of the optical cavity mode and sending the transmitted signal to a high-bandwidth (8~GHz) photoreceiver, the output of which is sent to a real-time electronic spectrum analyzer.  The use of high-bandwidth, low-noise photoreceivers in direct detection is essential due to the lack of low-noise optical amplifiers at 980~nm, in contrast to the 1550~nm band where near quantum-limited erbium-doped fiber amplifers (EDFAs) are readily available. For example, typical 10~GHz photoreceivers have a noise equivalent power $>20$~pW/$\sqrt{\textup{Hz}}$, so that at a frequency of 4~GHz, the expected noise power is $>1~$\textmu W (techniques to downshift a detected photocurrent to lower frequencies may reduce the expected noise power).  The typical thermal noise motional amplitude $x_{th}$ for one of our devices ($\approx~106$~fm at 293~K) yields an optical cavity mode frequency shift $\Delta f/2\pi=g_{0}/2\pi*x_{th}/x_{zpf}\approx0.002f_{o}/Q_{o}$, where $x_{zpf}\approx1.9$~fm is the zero-point motional amplitude, $f_{o}$ is the optical mode frequency, and $Q_{o}\approx10^5$ has been assumed.  The frequency modulation induced by the mechanical motion is transduced by the optical cavity into a fluctuating intensity, which for the same $Q_{o}$ will yield an intensity modulation that is less than 1$~\%$ of the input optical power.  This value must be larger than the expected noise power to yield a signal-to-noise level greater than unity in the mechanical mode spectroscopy. Alternately, homodyne detection~\cite{ref:Schliesser_NJP} can be used to increase signal levels above the detector noise floor.

Cryogenic measurements of nanophotonic devices using optical FTWs have previously been performed in systems such as a continuous-flow, liquid He-4 cryostat with the sample in vacuum~\cite{ref:Srinivasan14,ref:Chan_Painter_ground_state,ref:Liu_yuxiang_wlc} and a He-4 exchange gas cryostat~\cite{ref:Riviere_Kippenberg_cryostat}.  Here, we use a system similar to that in \cite{ref:Srinivasan14}, where low-temperature-compatible, piezo slip-stick positioners provide sample motion in the $x-y-z$ directions, a piezo slip-stick positioner provides adjustment of the out-of-plane angle of the fiber taper waveguide with respect to the sample, and a piezo flexure stage provides fine translation of the fiber taper in the $x-y-z$ directions.  A significant difference in the current setup is that cooling is provided in a closed-cycle fashion through a He compressor and Gifford-McMahon cryocooler. Improved thermal links with respect to those used in Ref.~\cite{ref:Srinivasan14} enable both the sample and fiber taper temperature to reach $\approx~$4.2~K, as measured by calibrated Si diodes mounted in adapter plates directly underneath the sample and fiber.



\subsection{Optomechanical Sideband Spectroscopy}
\label{sec:OM_sideband_spectroscopy}

One phenomenon observable in sideband-resolved devices is the optomechanical analog to electromagnetically induced transparency (EIT) and absorption (EIA)~\cite{ref:Weis_Kippenberg,ref:safavi-naeini4,ref:Agrawal_Huang_EIT}.  Here, the presence of a strong control field, appropriately detuned from the optical cavity, influences the transmission spectrum of a weak probe field swept in frequency across the optical mode.  It results from the interference of probe photons with photons from the control laser that are scattered by the mechanical resonator.  Such interference requires the probe photons and scattered photons to be phase coherent, which effectively shows that the optical field and mechanical mode can interact coherently.  The interference can be constructive or destructive, depending on which sideband photons interfere with the probe.

In particular, pumping the optical cavity with a red-detuned control beam creates a situation in which destructive interference takes place, leading to the observation of EIT. The EIT signal is a manifestation of a process in which an injected optical signal resonant with the cavity is converted into coherent mechanical motion, then reconverted into the optical domain, and so forth in a cycle~\cite{ref:safavi-naeini4,ref:Weis_Kippenberg}. This photon-phonon translation scheme~\cite{ref:safavi-naeini3}, occurs at a rate given by the pump-enhanced optomechanical coupling rate $G=g_0\sqrt{N}$, where $N$ is the number of control beam photons injected into the cavity; thus $G$ can be controlled via the control beam power. A number of optical signal processing functions have been proposed~\cite{ref:safavi-naeini3} and implemented~\cite{ref:safavi-naeini4,ref:Hill_Painter_WLC_Nat_Comm,ref:Liu_yuxiang_wlc} based on the photon-phonon translation (PPT) concept. Central to PPT is the optomechanical cooperativity parameter, $C=4G^2/\kappa\gamma_m$, essentially the ratio of the rate at which optical and mechanical energy are coherently exchanged (represented by the pump-enhanced optomechanical coupling rate $G$) over the loss rates of the mechanical and optical resonances ($\gamma_m$ and $\kappa$, respectively). Generally, $C\gg1$ is desired, meaning that optomechanical transduction happens at a considerably faster rate than energy is lost by the system.

Apart from demonstrating a manifestation of coherent interaction between the optical and mechanical degrees of freedom, EIT measurements allow the estimation of important system parameters, particularly relating to the photon-phonon translation process. The cooperativity $C$ can be directly obtained independently of the knowledge of the intracavity photon population or the vacuum optomechanical coupling $g_0$. In addition, the intrinsic mechanical and total optical linewidths can be obtained independently from the optical and mechanical spectroscopic methods described in Sec.~\ref{sec:OM_spectroscopy}~above. This capability can be very useful in situations where the apparent optical and mechanical linewidths are modified by the control beam, as discussed below.

In the EIT measurement setup, shown in Fig.~\ref{fig:Figure1}, light from the 980~nm tunable laser corresponds to the control field, and a probe signal is derived from it by modulation with an electro-optic phase modulator (EOPM). This produces out-of-phase blue and red sidebands at frequencies $\pm\Delta_\text{pc}$ away from the control beam frequency $\omega_c$. As shown Fig.~\ref{fig:Figure1}, the EOPM is driven by port 1 of a vector network analyzer (VNA), so that the probe-control beam detuning $\Delta_{\text{pc}}$ can be swept. For small phase modulation index $\beta$, the optical signal fed into the FTW can be represented by
\begin{equation}
E_\text{in} = e^{i\omega_ct}+\frac{\beta}{2}\left[ e^{i(\omega_c+\Delta_\text{pc})t} + e^{i(\omega_c-\Delta_\text{pc})t} \right].
\end{equation}
After the cavity, which has a transmission transfer function $t(\omega)=|t(\omega)|e^{i\phi(\omega)}$, this becomes

\begin{eqnarray}
E_\text{out} = e^{i\omega_ct} \left\{ t(\omega_c) + \frac{\beta}{2} \left[ t(\omega_c+\Delta_\text{pc}) e^{i\Delta_\text{pc}t} + \right. \right. \nonumber \\ \left. \left. t(\omega_c-\Delta_\text{pc}) e^{-i\Delta_\text{pc}t} \right] \right\},
\end{eqnarray}

This signal is then detected, yielding a photocurrent proportional to $|E_\text{out}|^2$. The photocurrent component $I_{\Delta_\text{pc}}$ oscillating at $\Delta_\text{pc}$ is
\begin{eqnarray}
I_{\Delta_\text{pc}}\propto\left\{ \left|t\left(\omega_c-\Delta_\text{pc}\right)\right|\cos \left(\Delta_\text{pc}t +\phi_+ \right)+ \right. \nonumber \\\left.\left|t\left(\omega_c+\Delta_\text{pc}\right)\right|\cos\left( -\Delta_\text{pc}t +\phi_- \right) \right\},
\end{eqnarray}
where $\phi_\pm=\angle t\left(\omega_c\right)-\angle t\left(\omega_c\pm\Delta_\text{pc}\right)$ ($\angle t$ is the phase of $t$). This can be expanded into an in-phase ($I$) and a quadrature ($Q$) component, $
I_{\Delta_\text{pc}}\propto\left\{ I\cdot\cos(\Delta_\text{pc}t) + Q\cdot\sin(\Delta_\text{pc}t)  \right\}$,
with
\begin{equation}
\label{eq:I}
I = \left|t\left(-\Delta_\text{pc}\right)\right|\cos(\phi_-)+\left|t\left(\Delta_\text{pc}\right)\right|\cos(\phi_+)
\end{equation}
and
\begin{equation}
\label{eq:Q}
Q = \left|t\left(-\Delta_\text{pc}\right)\right|\cos(\phi_-)-\left|t\left(\Delta_\text{pc}\right)\right|\cos(\phi_+).
\end{equation}
The $|S_{21}|$ parameter displayed by the VNA is such that $|S_{21}|\propto\sqrt{I^2+Q^2}$ and $\angle|S_{21}|=\tan^{-1}Q/I$. In practice, the sidebands acquire a relative phase while propagating down the fiber sections from the modulator to the cavity, and then from the cavity to the detector. To take that into account, we substitute $\phi_+\to\phi_++\theta$, and use $\theta$ as a fit parameter. In an ideal situation in which a phase modulator is used and no relative phase is acquired in the fiber, then $\theta=\pi$, so that the two sidebands are $180^\circ$ out-of-phase.

\begin{figure}[!t]
\centerline{\includegraphics[width=\linewidth]{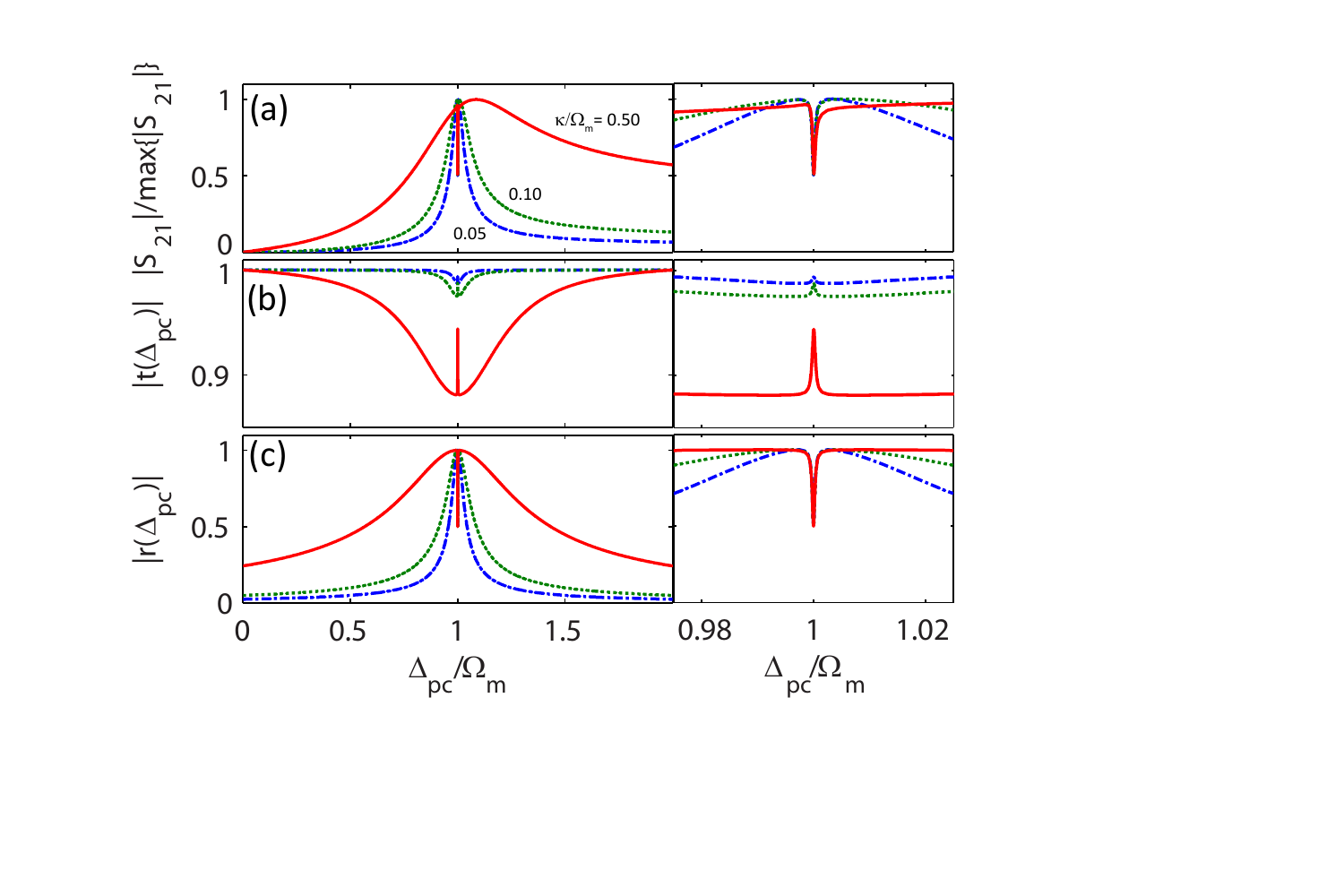}}\caption{Simulated vector network analyzer $|S_{21}|$ as a function of the signal-cavity detuning $\Delta_\text{pc}$  for a red-detuned control beam with $\Delta_\text{oc}=\Omega_m$, $Q_m=1000$, $\kappa_e = 0.25\cdot\kappa$, $\theta=\pi$, $C=1$ and varying sideband resolution $\kappa/\Omega_m$. (b) and (c) show corresponding transmission and reflection amplitude spectra, normalized to their maximum values. Panels to the right show the same curves, over a narrow frequency range around the EIT dip. In all plots, curves for $\kappa/\Omega_m = 0.5, 0.1 $~and $0.05$ are plotted with continuous, dotted and dash-dotted lines respectively.}\label{Fig:EIT_sims}
\end{figure}

For a red-detuned control field, the optomechanical cavity transmission coefficient $t\left(\Delta_{\text{pc}}\right)$, as a function of the control-probe detuning $\Delta_\text{pc}$ is given by
\begin{equation}
\label{eq:t}
t(\Delta_{\text{pc}}) = 1 -\frac{\kappa_{e}/2}{i(\Delta_{\text{oc}}-\Delta_{\text{pc}})+\kappa/2+\frac{G^2}{i(\Omega_{\text{m}}-\Delta_{\text{pc}})+\gamma_{\text{m}}/2}},
\end{equation}
where $\kappa_e$ is the fiber-cavity coupling rate, $\kappa$ is the optical cavity decay rate, $\Delta_\text{oc}$ the cavity-control beam detuning, $\Omega_m$ the mechanical frequency, $\gamma_m$ the intrinsic mechanical damping and $G=g_0\sqrt{N}$ the pump-enhanced optomechanical coupling, with $N$ the intracavity photon number. The cavity transmission spectrum in eq.~(\ref{eq:t}) can be plugged into eqs.~(\ref{eq:I}) and (\ref{eq:Q}) to give the expected VNA $|S_{21}|$ parameter (scaled by a constant factor), which can be used for fitting experimental data. Fig.~\ref{Fig:EIT_sims}(a) shows an example $|S_{21}|$ obtained for an optomechanical cavity with $\Delta_\text{oc}=\Omega_m$, $Q_m=1000$, $\kappa_e = 0.25\cdot\kappa$, $\theta=\pi$, $C=1$ and $\kappa/\Omega_m \in [0.05,0.1,0.5]$. Figs.~\ref{Fig:EIT_sims}(b) and (c) show the amplitude of the corresponding cavity transmission and reflection coefficients. Panels to the right show a blow-up of the corresponding graphs in the vicinity of the EIT dip.

We note that both $I$ and $Q$ in eqs.~(\ref{eq:I}) and~(\ref{eq:Q}) combine contributions from both sidebands of the modulated signal, $t(\pm\Delta_\text{pc})$ for the Stokes and Anti-Stokes, respectively. For sufficient sideband resolution, $t(\omega_c)\approx t(\omega_c-\Delta_\text{pc})\approx1$ for red sideband detuning. In this case, for $\theta=\pi$ (corresponding, as above, to phase modulation and no relative phase gain through the fibers), it can be shown that $|S_{21}|$ approximates $|r(\omega_c+\Delta_\text{pc})|$, where $r=1-t$ is the cavity reflectivity coefficient. This is apparent in a comparison between $S_{21}$ in Fig.~\ref{Fig:EIT_sims}(a) and the corresponding reflection coefficients in Fig.~\ref{Fig:EIT_sims}(c), for varying levels of sideband resolution $\kappa/\Omega_m$: for a less well-resolved cavity (large $\kappa/\Omega_m$), the $|S_{21}|$ becomes increasingly more skewed with respect to $|r|$.

While some of the fitting parameters can be obtained independently from other measurements, we note that in certain situations, extraction with the VNA method can be advantageous, for practical reasons. The optical cavity external and total losses $\kappa_e$ and $\kappa$ can be obtained from a power transmission spectrum $T(\lambda)$, obtained by scanning the laser wavelength $\lambda$ across the cavity and fitting with eq.~(\ref{eq:t}) with $G=0$ (note that in these measurements, the scanned laser acts as a signal, not as a control for the optomechanical interaction). The total loss rate $\kappa$ incorporates both the external coupling rate and intrinsic (absorption and parasitic) losses: $\kappa=\kappa_e+\kappa_i$. $\kappa$ and $\kappa_e$ are connected by the transmission at cavity center $T_0$, $\kappa_e=\kappa\cdot\sqrt{1-T_0}$. If the cavity displays significant nonlinear absorption, it is expected that $\kappa$ will change with the injected control beam power, and so extracting $\kappa$ by using it as a fit parameter in eq.~(\ref{eq:t}) could be useful in these conditions. Likewise, the intrinsic mechanical linewidth $\gamma_m$ can be obtained from the mechanical thermal motion spectrum. In practice, however, it is often difficult to measure such spectra with sufficient signal-to-noise ratio with a low-power control beam. Particularly in sideband-resolved systems, the control beam power needs to be sufficiently low that dynamic back-action effects do not significantly modify the observed mechanical frequency and linewidth from their intrinsic values (at telecom wavelengths, this issue can be circumvented by amplifying the signal with an Erbium-doped fiber amplifier before detection). If such condition cannot be achieved unambiguously, therefore, extracting $\gamma_m$ by using it as a fit parameter for the VNA data can be a viable option.

\section{Fabrication and Improved Optical Performance}
\label{sec:Fabrication_higher_Q}
\begin{figure*}[!t]
\centering
\includegraphics[width=\textwidth]{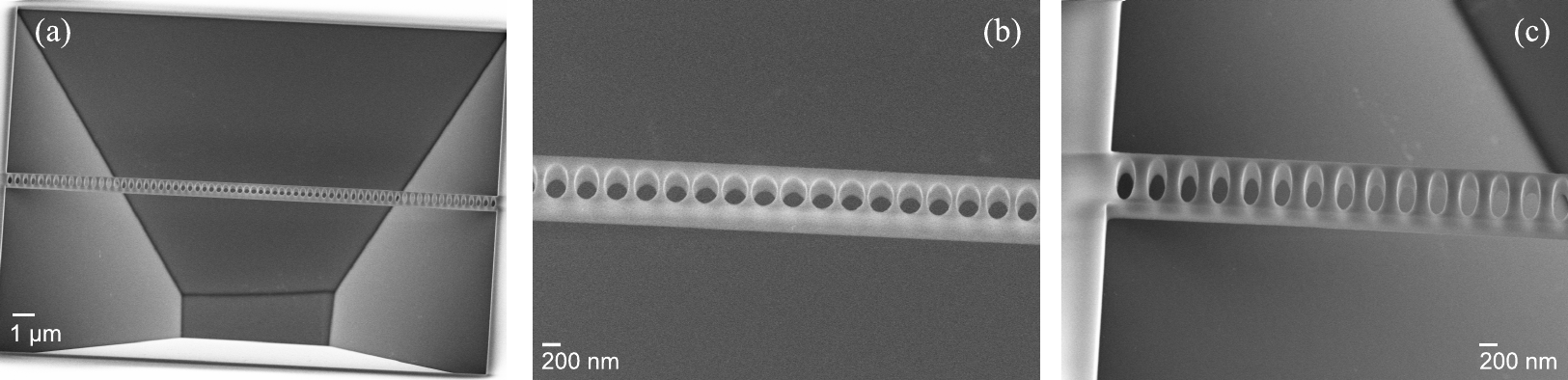}
\caption{SEMs of an etched and released optomechanical nanobeam, with wide-angle view in (a), zoomed-in view of center in (b), and zoomed-in vew of end mirror section in (c).}
\label{SEMs}
\hfil
\end{figure*}

Optomechanical crystal nanobeams were fabricated in 350~nm thick stoichiometric Si$_3$N$_4$ deposited via low-pressure chemical-vapor deposition on a plain Si substrate (tensile stress $\approx$~800~MPa). An array of device designs was patterned via electron-beam (E-beam) lithography in positive E-beam resist and developed in hexyl acetate.  The Si$_3$N$_4$ was etched by a CF$_4$/CHF$_3$ (4:1) reactive ion etch at 1.33~Pa chamber pressure. This etch recipe produces sidewall angles within 3$^\circ$ of vertical, as measured in cross-sections under a scanning-electron microscope (SEM).  According to finite element method simulations of the optical mode, this sidewall angle should have a minimal effect on device performance.  After the etch and removal of the E-beam resist, the devices were released in $45~\%$ KOH at $75~^\circ$C for about 15~min followed by a 5~min dip in 1:4~H$_2$O:HCl to remove precipitated residue from the KOH release. The nanobeams were robust enough that they could be simply N$_2$ blow-dried. SEM images of an example device are shown in Fig.~\ref{SEMs}.

\subsection{Development}
The multitude of surfaces in an OMC makes it important that the sidewalls be smooth in order to minimize optical scattering loss.  One source of sidewall roughness is the line edge roughness in the lithography.   Previous works have observed that the temperature of the development step for E-beam resists strongly affects the line edge roughness of E-beam patterns, as observed under SEM \cite{wang2007low,ocola2006effect}.

\begin{figure*}[!t]
\centering
\includegraphics[width=\textwidth]{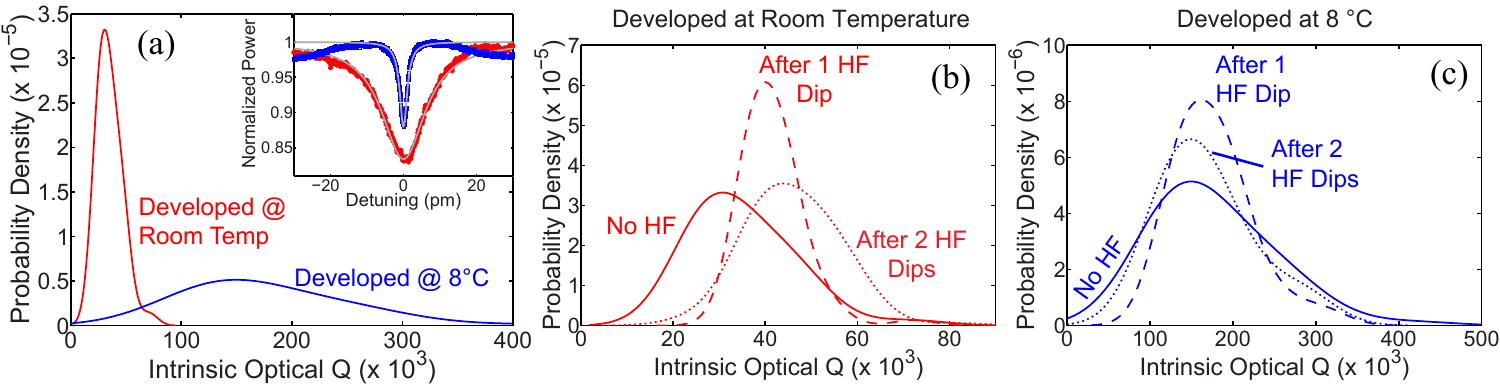}
\caption{(a)~The estimated probability density distribution of the intrinsic optical $Q$ in the fourteen devices measured from a room-temperature-developed sample and the fifteen devices measured from a sample developed at $8~^\circ$C. The devices in both samples were the same set of designs and the same E-beam exposure. The probability density is weighted by the uncertainty in the optical resonance fits. (inset) Data and fits of the highest-$Q$ resonances in the room-temperature-developed sample (red) and the cold-developed sample (blue). The room-temperature-developed device had a loaded $Q_o$ of $\left(6.8\pm0.1\right)\times10^4$, corresponding to an intrinsic $Q_o$ of $\left(7.1\pm0.1\right)\times10^4$.  The highest measured loaded $Q_o$ in the cold-developed sample was $\left(4.0\pm0.3\right)\times10^5$, which corresponds to an intrinsic $Q_o$ of $\left(4.1\pm0.3\right)\times10^5$. $Q$s are determined by a nonlinear least squares fit of the data. The uncertainties are given by the $95~\%$ confidence intervals of the fit. (b)~Estimated probability density distributions of the intrinsic optical $Q$s before HF treatment and after HF treatments in the room-temperature-developed sample. The sample was measured after one 2~min dip in 50:1 HF and again after a second dip. (c)~ Estimated probability density distributions of the intrinsic optical $Q$s before HF treatment and after HF treatments in the sample developed at $8~^\circ$C. The sample was measured after one 2~min dip in 50:1 HF and again after a second dip.}
\label{Qdistr}
\hfil
\end{figure*}

Here we quantify the difference between cold- and room-temperature-developed E-beam resist by measuring the optical $Q$.  We exposed several sets of the same device designs and dosages, but developed one set at room temperature and the other set at $8~^\circ$C.  For the room-temperature developed sample, the highest measured loaded $Q_o$ was $\left(6.8\pm0.1\right)\times10^4$, corresponding to an intrinsic $Q_o$ of $\left(7.1\pm0.1\right)\times10^4$.  (The OMCs demonstrated in Ref. \cite{ref:Davanco_nanobeam_OMC} were fabricated via a similar process, resulting in $Q$s as high as $1.2\times10^5$, with the average $Q$s around $7.5\times10^4$.) The highest measured loaded $Q_o$ in the cold-developed sample was $\left(4.0\pm0.3\right)\times10^5$, which corresponds to an intrinsic $Q_o$ of $\left(4.1\pm0.3\right)\times10^5$ (Fig.~\ref{Qdistr}(a) inset). $Q$s are determined by a nonlinear least squares fit of the data. The uncertainties are given by the $95~\%$ confidence intervals of the fit.

An estimated probability density function based on all the measured $Q$s and weighted by their uncertainties is shown in Fig.~\ref{Qdistr}(a). Essentially, these functions are constructed by summing normal distributions centered at the optical $Q$ of each of the devices, where the distribution variance of each is the uncertainty in $Q$. The resulting summed function is then smoothed and normalized such that its integral is one. For the room-temperature-developed sample, the optical $Q$s are tightly clustered around $2.9\times10^4$, while the $Q$s of the cold-developed sample are more broadly distributed, centered around $1.3\times10^5$.  The tight clustering and lower $Q$ of the room-temperature-developed devices suggests they are all limited by the same upper bound on the optical $Q$, likely surface roughness, whereas the wide distribution of the cold-developed sample suggests that the optical $Q$ is not dominated by the same factor for all devices; the variation is likely due to the variation in the exposed optomechanical crystal design.


\subsection{Surface Treatment}
In addition to sidewall roughness, it is possible that surface absorption plays a significant role in the optical loss.  In LPCVD Si$_3$N$_4$, oxygen and carbon surface contamination is very common \cite{guillermo2014evidence}.  To test the significance of surface absorption, we etched the released nanobeams in 50:1 HF for 2~min, removing about 1~nm of material from all surfaces, as determined by ellipsometric measurements of Si$_3$N$_4$ film thickness as a function of etch time when exposed to 50:1 HF.  We then characterized the subsequent optical performance of the optomechanical crystals.  The optical resonant wavelength decreased by around 3~nm to 4~nm after each HF dip.  An estimated probability density function of the intrinsic $Q_o$s before and after one and two HF dips is shown in Fig.~\ref{Qdistr}(b) and (c). In the room-temperature developed devices, the most probable intrinsic $Q_o$ increases by less than a factor of two after two HF dips, while the cold-developed devices stay about the same.  It is likely that the HF treatments somewhat reduced the surface roughness in the room-temperature-developed samples, while, in the smoother cold-developed samples, the dominant effect was changing the photonic crystal dimensions.

Thus, the dilute HF etching shown here does not appreciably modify the $Q_o$ of cold-developed samples, indicating that either surface absorption is not a dominant loss mechanism for these devices, or that the absorbing layer has not been effectively removed in this process.  We did see some improvement in $Q_o$ for room-temperature developed samples, which is consistent with an improvement in surface roughness.  For longer dip times or stronger HF concentrations, the HF dip should be taken into consideration when designing the cavity geometry.

In a couple of devices, we also characterized the optical resonance wavelength change with respect to input power with the coupling distance fixed.  This type of measurement  indicates any non-linear optical absorption and can be used to determine the linear absorption of a material \cite{borselli2007accurate}. Non-linear absorption would result in a change in the extinction ratio with respect to the optical power in the cavity.  Both before and after HF treatment, the extinction ratio varied less than 1~$\%$ and with no discernible trend for resonant wavelength changes of up to 10~pm. This small variation indicates insignificant levels of non-linear absorption in the Si$_3$N$_4$ film.

Because surface effects can also degrade mechanical quality factors, we also tested the effectiveness of the HF treatment for improving $Q_m$. We measured the mechanical quality factor of the approximately 3.8~GHz breathing mode with low input optical power to minimize the effects of dynamical backaction.  There was no significant change in $Q_m$ as a result of HF treatment in the devices we measured, implying that either surface loss is insignificant in the breathing mode or dipping in HF does not effectively remove mechanical surface loss mechanisms.



\section{Temperature-Dependent Measurements}
\label{sec:Temp_dep_measurements}
There have been several demonstrations of mechanical damping decreasing as temperature decreases in Si$_3$N$_4$ flexural modes (mechanical frequencies less than 1~GHz) \cite{ref:Zwickl_Harris,southworth2009stress,faust2014signatures,yasumura2000quality}.  These works point to two-level systems and atomic tunneling in glasses as limits on the mechanical $Q$, but it is not clear how the significance of these effects would change for a high-frequency bulk mode, like the breathing mode of our optomechanical crystals. The breathing mode (around 3.6~GHz) of crystalline Si optomechanical nanobeams has been measured at low temperatures, with a mechanical $Q$ on the order of $10^6$ measured at 10~K \cite{ref:chan_optimized_OMC}, and a mechanical $Q$ inferred to be $9\times10^6$ at 10~mK \cite{ref:Meenehan_Painter_pra}.  Thus, there is some indication that very high mechanical $Q$s in breathing modes can be achieved at low temperatures.



\begin{figure*}[!t]
\centering
\includegraphics[width=\textwidth]{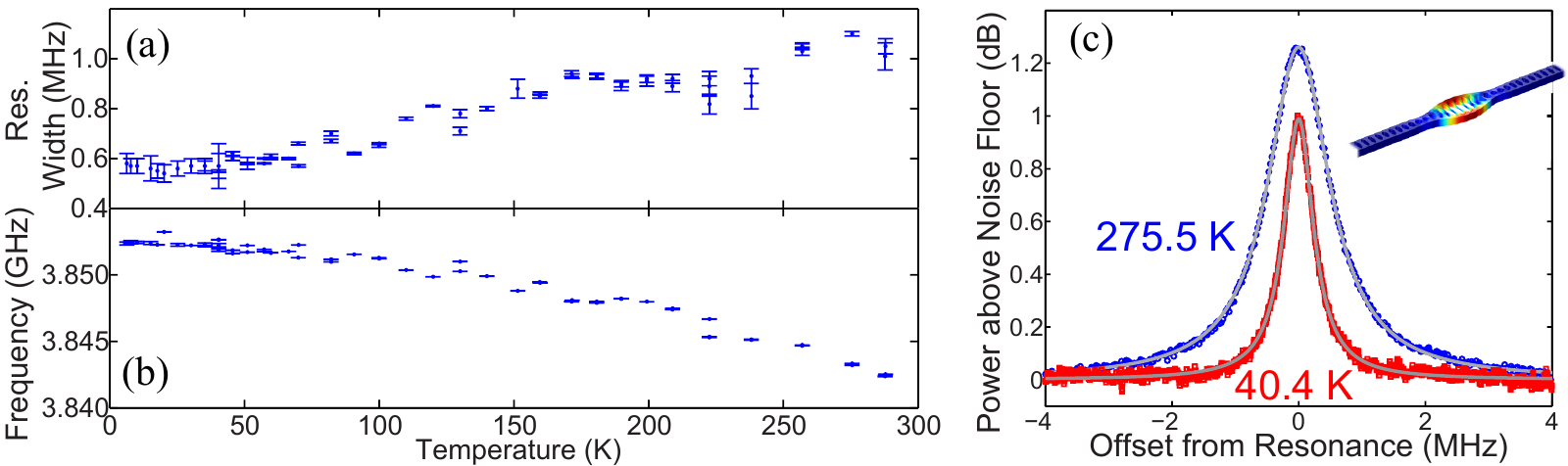}
\caption{(a)~Measured mechanical resonance width as a function of temperature. Resonance frequencies and widths are determined by a nonlinear least squares fit of the data. The uncertainties are given by the $95~\%$ confidence intervals of the fit. (b)~Measured mechanical breathing mode frequency as a function of temperature. Error bars come from the uncertainty in the fit of the mechanical resonance. (c)~Examples of mechanical resonances and their Lorentzian fits. The resonance measured at 275.5~K had a mechanical $Q$ of $3500\pm30$, while the resonance measured at 40.4~K had a mechanical $Q$ of $7070\pm60$.  The input optical power for these two measurements was about the same to within 10~$\%$. (inset) A finite element method model of the breathing mode.}
\label{TempMech}
\hfil
\end{figure*}

To characterize the significance of temperature in the mechanical performance of our Si$_3$N$_4$ optomechanical crystals, we measured the mechanical $Q$ of an optomechanical crystal from 6~K to room temperature in a temperature-controlled, closed-cycle, Gifford-McMahon cryostat.  The mechanical modes of the nanobeam were detected as described in Sec.~\ref{sec:OM_spectroscopy}. The optical power was kept low to minimize the effects of dynamical back-action on the mechanical linewidth.  In this device, the first-order breathing mode (Fig.~\ref{TempMech}(c) inset) had a mechanical frequency of about 3.8~GHz.  The linewidth change with respect to temperature is shown in Fig.~\ref{TempMech}.  The linewidth at room temperature was 1.05~MHz~$\pm0.03$~MHz, corresponding to a $Q_m$ of $3700\pm100$.  The quality factor increased as the temperature decreased, leveling out at about 70~K, at which point the linewidth was 0.58~MHz~$\pm0.01$~MHz $\left(Q_m=6700\pm100\right)$. $Q_m$s are extracted from a nonlinear least squares fit of the data. The uncertainties are given by the $95~\%$ confidence intervals of the fit.

This reduction in damping by almost half is comparable to the magnitude of improvement in the flexural modes of Ref. \cite{faust2014signatures}, but we do not see the same two-level-system dissipation peak around 50~K. We do see some indication of a similar, small dissipation peak around 170~K, which could be the result of a thermally-activated relaxation due to hydrogen defects in the Si$_3$N$_4$.  Other factors that might be limiting $Q_m$ include clamping loss due to incomplete mode confinement in the fabricated optomechanical crystal and localized defect states \cite{unterreithmeier2010damping}.  Thermoelastic damping and Akhieser losses are likely not the dominant factors limiting $Q_m$, as the simulated and calculated \cite{ghaffari2013quantum} $Q_m$ limits due to these effects are at least an order of magnitude greater than the measured mechanical $Q$s. As of yet, more study is required to determine how to mitigate the remaining mechanical loss mechanisms, but the $f_m$$\times$$Q_m$ product that our OMC achieves at temperatures less than 50~K ($\left(2.59\pm0.04\right)\times10^{13}$~Hz) is on par with the highest previously demonstrated in Si$_3$N$_4$ mechanical resonators~\cite{wilson2009cavity,purdy2012cavity}.  Interestingly, these previous demonstrations were in very different geometries (e.g., MHz frequency membrane modes).  Perhaps more importantly, this $f_m$$\times$$Q_m$ product is relevant in the context of understanding how decoupled the mechanical mode of interest can be from the thermal environment it is contacting~\cite{ref:Aspelmeyer_Kippenberg_Marquardt_Review}.  The rate at which thermal phonons can be coupled into the mechanical resonator is given by $\gamma_{m}n_{th}$, where $n_{th}=k_{b}T/\hbar\Omega_{m}$ is the number of thermal phonons at the mode frequency $\Omega_{m}$ for a temperature $T$, and $k_{b}$ is Boltzmann's constant.  The quantity $\frac{\Omega_{m}}{\gamma_{m}n_{th}}=h\frac{f_mQ_m}{k_{b}T}$ thus represents the number of coherent oscillations that can take place before this thermal decoherence sets in, and is $>100$ for our devices.  This environmental decoupling would be of particular importance if these systems can be cooled to the quantum ground state.

\section{EIT Measurements at Cryogenic Temperatures}

\begin{figure}[!t]
\centerline{\includegraphics[width=\linewidth]{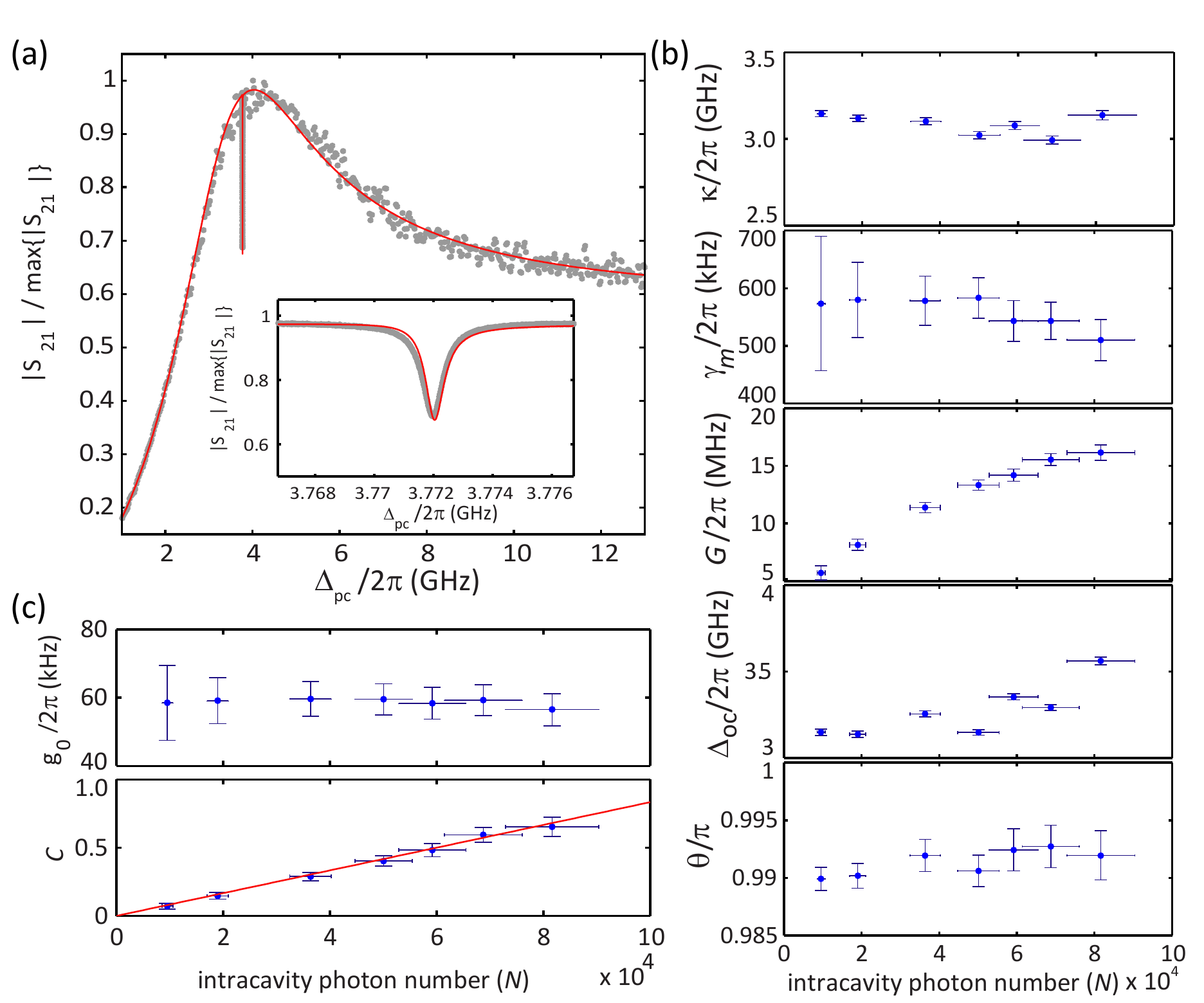}}\caption{Vector network analyzer $|S_{21}|$ scan of $\SiN$ OMC for a red-detuned control beam. Grey dots: data; red line: fit. The frequency $\Delta_\text{pc}$ is the spacing between the control beam and modulation sidebands (probe signal). Inset: blow-up of the frequency range where EIT is observed. (b) Optical cavity linewidth $\kappa$, intrinsic mechanical linewidth $\gamma_m$, pump-enhanced optomechanical coupling $G$, control beam-cavity detuning $\Delta_\text{oc}$ and relative sideband phase $\theta$ as functions of the intracavity photon number $N$, all obtained from fits as in (a). (c) Optomechanical coupling rate $g_0$ and cooperativity $C$ as functions of $N$. Red line is a linear fit.
} \label{Fig:EIT}
\end{figure}

Fig.~\ref{Fig:EIT}(a) shows representative experimental VNA data for a nanobeam OMC tested at 30 K (grey dots) under red-detuned pumping, together with a fitted curved obtained using $\kappa$, $\gamma_m$, $C$, $\Delta_\text{oc}$ and $\theta$ as parameters. The experimental data displays a narrow EIT dip at $\Delta_\text{pc}\approx3.77$~GHz, shown in the inset. In order to correct for the frequency response of the photoreceiver, the raw $|S_{21}|$ was normalized by a background $|S_{21}|$ signal taken with a far-detuned pump with polarization tuned to maximize amplitude modulation at the electro-optic phase modulator. Because the optical cavity linewidth is wide compared to the photoreceiver bandwidth, this normalization procedure is important and must be done with care in order for fits to be properly done. It is apparent that the cavity model is able to describe the data well over the entire displayed frequency range, including the narrow range around the EIT dip (inset). In this fit, the ratio $\kappa_e/\kappa=0.24$ was obtained from a fit to the cavity transmission spectrum, so that $\kappa_e$ was not a fit parameter to the $|S_{21}|$ data. In addition, the mechanical frequency $\Omega_m$ was taken at the minimum of the EIT dip. The fit procedure was repeated for varying red-detuned pumps at various powers, and the extracted cavity parameters are shown in Fig.~\ref{Fig:EIT}(b) as a function of the intracavity photon number $N$. The error bars correspond to $95~\%$ fit confidence intervals, and are due to noise in the experimental data. $N$ is calculated with the expression
\begin{equation}
N = \frac{1}{\hbar\omega_{o}}\eta\Delta TQ_{i}\left(\frac{P_{\text{in}}}{\omega_o}\right)\frac{1}{1+(\frac{\Delta_{\text{oc}}}{\kappa/2})^2},
\label{eq:Seq2}
\end{equation}
using parameters from the $|S_{21}|$ and transmission spectrum fits. Here, $\hbar=h/2\pi$ ($h$ is Planck's constant), $\eta$ is the FTW coupling efficiency, $\Delta T$ is the depth of the optical resonance in the transmission spectrum, $Q_i$ the intrinsic optical $Q$, and $P_{\text {in}}$ is the optical power at the FTW input. The intrinsic $Q$ was obtained as $Q_i= 2Q /(1+\sqrt{1-\Delta T})$.  Within our error bars, the intrinsic optical and mechanical damping rates are constant with pump power, and the phase between the sidebands is very close to $\pi$. The average mechanical linewidth is such that the mechanical quality factor at 30~K extracted here is approximately 2$\times$ that at room temperature, consistent with the data from Sec.~\ref{sec:Temp_dep_measurements}. The scatter in $\Delta_\text{oc}$ reflects the manual selection of the control beam-cavity detuning at each pump power tested. The pump-enhanced optomechanical coupling $G$ follows the expected $\sqrt{N}$ shape, which is reflected in the linear trend observed for the cooperativity $C=4G^2/\kappa\gamma_m$ shown in Fig.~\ref{Fig:EIT}(c). The cooperativity here is approximately twice that observed in~\cite{ref:Davanco_nanobeam_OMC}, likely due to the smaller $\gamma_m$ and $\kappa$, and despite the lower optomechanical coupling rate $g_0=G/\sqrt{N}$, also shown in Fig.~\ref{Fig:EIT}(c). We note that $N$ can be difficult to accurately measure as it relies upon good knowledge of the pump-cavity detuning and all optical losses in this system.  For this reason, other methods to assess $g_{0}$ that are independent of $N$~\cite{ref:Gorodetsky_Kippenberg_OM} may be preferable, and could be used together with extracted cooperativity values to determine the intracavity photon population.

\section{Discussion}

We have presented measurements of $\SiN$ nanobeam optomechanical crystals in which a 3.8~GHz mechanical breathing mode is coupled to a 980~nm optical mode.  Compared to our earlier work~\cite{ref:Davanco_nanobeam_OMC}, we have been able to increase optical quality factors (both the average and maximum value across many devices) by as much as a factor of 4, with a highest $Q_o=4.1{\times}10^5$ measured.  We have also looked at the effect of weak HF etching on both optical and mechanical $Q$, to assess whether it is effective at removing potential surface loss mechanisms.  Finally, we have performed temperature-dependent measurements from 6~K to 300~K, and have seen an improvement in mechanical $Q$ by about a factor of 2.

At a mode temperature of 6~K, the phonon occupation number for the 3.8~GHz breathing mode (without laser cooling) is calculated to be  $<n>=k_{b}T/\hbar\Omega_m\approx$~33. For applications such as laser cooling of the mechanical mode to lower occupation levels, much higher cooperativity values than those demonstrated here ($C~\approx~$0.6) must be achieved.  This can be done through an increased intracavity photon number $N$, higher optomechanical coupling rate $g_{0}$, lower mechanical dissipation rate $\gamma_{m}$, and lower optical decay rate $\kappa$.  $N$ is currently limited by thermo-optic dispersion, where heating of the optical cavity (for example, due to absorption) results in a refractive index change and a shift in the optical cavity mode frequency.  While in devices like microcavity frequency combs, a so-called `soft thermal lock'~\cite{ref:del_haye_Kippenberg_comb} can be achieved to effectively lock the cavity detuning with respect to the input laser, we have not been able to reproduce a similar effect in the $\SiN$ nanobeam optomechanical crystals, where the detuning level (in units of number of optical cavity linewidths) is much larger, placing it at a thermally unstable point \cite{carmon2004dynamical}.  Significant increases in $N$ will either require reduced heating (e.g., lower absorption levels) or some mechanism to lock the laser with respect to the cavity in the presence of thermo-optic dispersion.

In this work, we have shown improved optical and mechanical quality factors with respect to those in Ref.~\cite{ref:Davanco_nanobeam_OMC}.  Further improvements in mechanical $Q$ will require a more detailed understanding of dissipation mechanisms at GHz frequencies in $\SiN$ (we note that $\approx$~10~MHz frequency devices fabricated using the same process as described here have $Q_{m}~\gtrsim~10^5$ at room temperature and under vacuum). Cryogenic cooling to even lower temperatures may minimize remaining potential dissipation mechanisms (e.g., two-level systems \cite{mohanty2002intrinsic}); other approaches to passivate $\SiN$ surfaces may also be considered.  Regarding optical losses, we have found an improved electron-beam lithography process to be key to the improvement shown in this work.  Further improvements in lithography (e.g., through better proximity effect correction) will be the subject of future work.

Finally, one challenge in working with $\SiN$ relative to materials like Si is its comparatively lower refractive index, which causes a significant reduction in the coupling rate $g_{0}$ for similarly designed optomechanical structures.  Slot mode geometries, such as the optomechanical crystals designed in Ref.~\cite{ref:Davanco_OMC}, are one approach to increasing $g_{0}$, while also enabling a host of multimode applications in which the mechanical mode is coupled to more than one optical resonance, or vice versa.

\section{Acknowledgements}

We thank Kerry Neal and Isaac Henslee from Montana Instruments for assistance with the cryogenic measurement system. K.E.G. acknowledges the National Research Council for her NIST/NRC postdoctoral fellowship support. This work was also partly supported by the DARPA MESO project. The identification of any commercial product or trade name is used to foster understanding.  Such identification does not imply recommendation or endorsement or by the National Institute of Standards and Technology, nor does it imply that the materials or equipment identified are necessarily the best available for the purpose.

\end{document}